\documentclass[journal=jacsat,manuscript=article]{achemso}
\setkeys{acs}{keywords = true}
\setkeys{acs}{articletitle=true}

\usepackage{soul}
\usepackage{xcolor}
\usepackage[version=3]{mhchem} 

\author{Ali K. Jahromi}
\affiliation[University of Central Florida]
{CREOL, The College of Optics and Photonics, University of Central Florida, Orlando, FL 32816, United States}
\author{Soroush Shabahang}
\affiliation[University of Central Florida]
{CREOL, The College of Optics and Photonics, University of Central Florida, Orlando, FL 32816, United States}
\author{H. Esat Kondakci}
\affiliation[University of Central Florida]
{CREOL, The College of Optics and Photonics, University of Central Florida, Orlando, FL 32816, United States}
\author{Seppo Orsilla}
\affiliation[Modulight]
{Modulight, Inc., Hermiankatu 22, FI-33720, Tampere, Finland}
\author{Petri Melanen}
\affiliation[Modulight]
{Modulight, Inc., Hermiankatu 22, FI-33720, Tampere, Finland}
\author{Ayman F. Abouraddy}
\affiliation[University of Central Florida]
{CREOL, The College of Optics and Photonics, University of Central Florida, Orlando, FL 32816, United States}

\email{raddy@creol.ucf.edu}

\title[An \textsf{achemso} demo]
  {Transparent Perfect Mirror}


\begin{document}

\begin{tocentry}

Some journals require a graphical entry for the Table of Contents.
This should be laid out ``print ready'' so that the sizing of the
text is correct.

Inside the \texttt{tocentry} environment, the font used is Helvetica
8\,pt, as required by \emph{Journal of the American Chemical
Society}.

The surrounding frame is 9\,cm by 3.5\,cm, which is the maximum
permitted for  \emph{Journal of the American Chemical Society}
graphical table of content entries. The box will not resize if the
content is too big: instead it will overflow the edge of the box.

This box and the associated title will always be printed on a
separate page at the end of the document.

\end{tocentry}

\begin{abstract}
  A mirror that reflects light fully and yet is transparent appears paradoxical. Current so-called transparent or `one-way’' mirrors are not perfectly reflective and thus can be distinguished from a standard mirror. Constructing a transparent `perfect’' mirror has profound implications for security, privacy, and camouflage. However, such a hypothetical device cannot be implemented in a passive structure. We demonstrate here a transparent perfect mirror in a non-Hermitian configuration: an active optical cavity where a certain pre-lasing gain extinguishes Poynting's vector at the device entrance. At this threshold, all vestiges of the cavity’s structural resonances are eliminated and the device presents spectrally flat unity-reflectivity, thus becoming indistinguishable from a perfect mirror when probed optically across the gain bandwidth. Nevertheless, the device is rendered transparent by virtue of persisting amplified transmission resonances. We confirm these predictions in two photonic realizations: a compact integrated active waveguide and a macroscopic all-optical-fiber system.
\end{abstract}

\begin{figure}[h]
\begin{center}
\includegraphics[width=16.7cm]{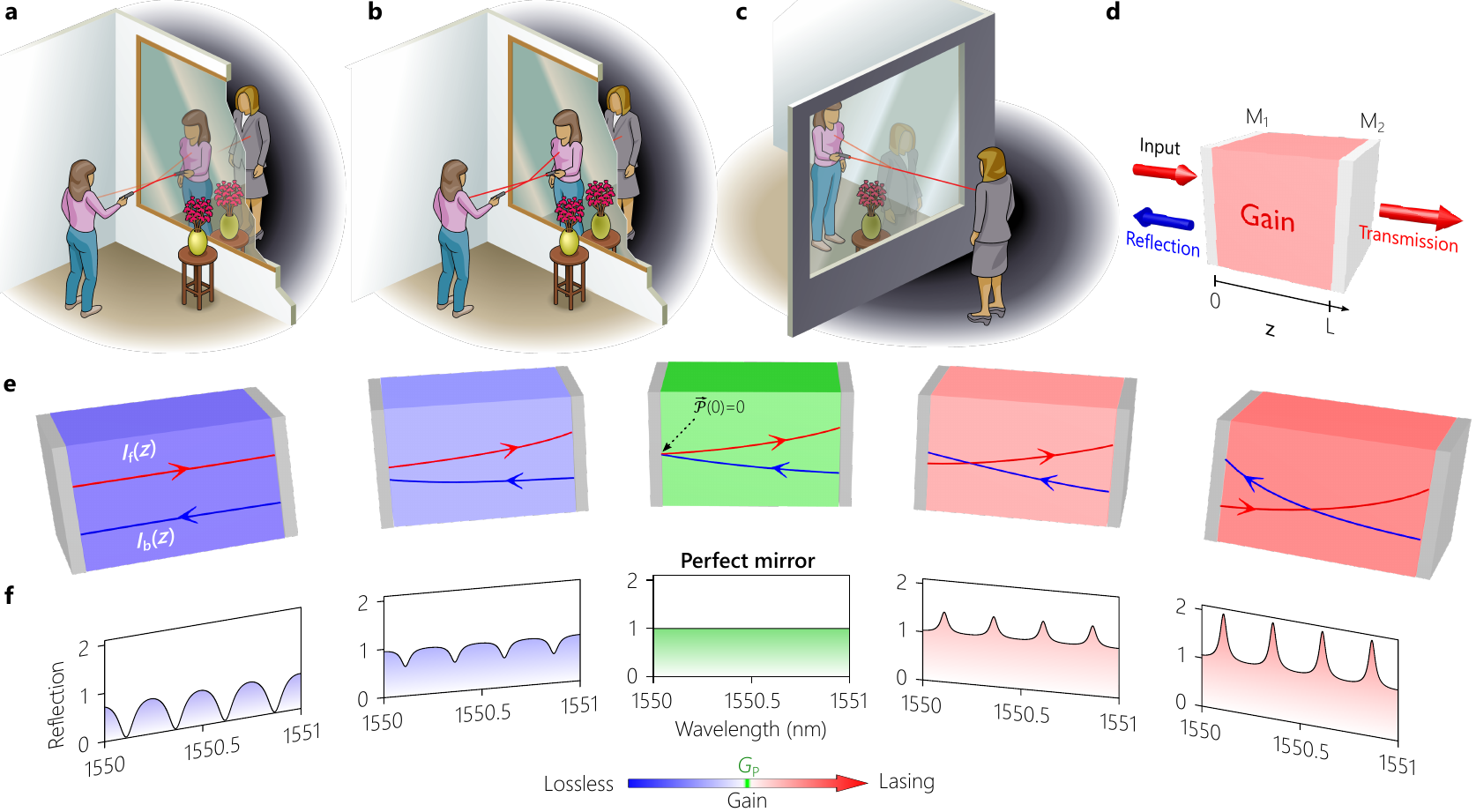}
\end{center}
\caption{(a) Concept of a transparent perfect mirror. (a) Illustration of a traditional `transparent mirror’'. Viewed from a brightly lit side, it appears reflective; the laser beam reflects only partially. (b) Illustration of a `transparent perfect mirror’'. Viewed from one side, it cannot be distinguished from a perfect mirror through optical interrogation. The laser beam reflects 100\%. (c) Viewed from the other side, the mirror appears transparent. (d) Schematic of an active cavity model formed of mirrors $\mathrm{M_1}$ and $\mathrm{M_2}$ with reflectivity $R_1$ and $R_2$, respectively, sandwiching a layer of thickness $L$
providing single-pass net-amplification $G$. (e) The forward and backward waves $I_\mathrm{f}(z)$ and $I_\mathrm{b}(z)$, respectively, along the planar cavity assuming $R_1=R_2=0.3$, $L=1.4\,\mathrm{mm}$ and the refractive index is 3.4. The single-pass net-amplification $G$ increases from one panel to the next; from left to right: lossless cavity $G=0\,\mathrm{dB}$, $G=2.3\,\mathrm{dB}$, Poynting's threshold $G=G_\mathrm{P}\approx2.\,\mathrm{dB}$, $G=2.85\,\mathrm{dB}$, $G=3.05\,\mathrm{dB}$ preceding lasing. Poynting's vector is extinguished at the cavity entrance when since. (f) Spectral reflection from the cavity for the values of $G$ in (e). At the transparent-perfect-mirror condition is satisfied and, signifying spectrally flat unity-reflectivity.
\label{fig:Concept}}
\end{figure}

Light transmits through a transparent window pane, whereas
a perfect mirror reflects light completely. Combining the two functions in the same device -- a so-called `one-way'’ or `see-through’' mirror \cite{Bloch1903, Bengoechea2004} –- can be useful in a variety of applications, such as secure surveillance, advertising, displays, and more recently in augmented reality devices and wearable technology \cite{Javidi2002, Barfield2016}. A `transparent mirror’' was proposed as early as 1903 that consists of a glass sheet provided with a partially reflective coating \cite{Bloch1903}. This device, however, is neither a perfect mirror nor transparent, but it can approximate the intended function when the ambient illumination on its two sides is different. This is commonly experienced when one looks through a window in a well-lit room at night. Viewed from the indoor bright-lit side, partial reflection from the window overwhelms the darkness outside and thus gives the impression of strong reflectivity, while appearing transparent from the dark outside. Therefore, any realization of a so-called transparent mirror is readily uncovered by monitoring the lack of complete reflection of an optical probe [Fig.~\ref{fig:Concept}(a)]. The transparent-mirror principle has undergone no fundamental change in the century since its inception.

We pose the following question: can one construct a `transparent perfect mirror’'? This is a mirror that reflects light completely and is indistinguishable from a perfect reflector – and yet transmits light. A passive implementation of this hypothetical device would involve a violation of energy conservation, which necessitates employing instead a non-Hermitian (non-conservative) arrangement; specifically, one provided with optical gain \cite{Siegman1986}. The sum of transmittance and reflectance from such a structure may indeed exceed unity, thereby supplying the energy deficit that allows for finite transmission of incident light while maintaining unity-reflectivity. Nevertheless, simply adding gain does not
guarantee perfect reflection. Rather, gain must be added in such a way as to provide no signatures of the underlying structure employed, and present instead spectrally flat unity-reflectivity.

Recent developments have demonstrated that non-Hermitian photonic structures -- in which optical gain and loss are judiciously distributed \cite{Regensburger12Nature} -- can indeed expand the range of realized optical functionalities in photonic devices when compared to their Hermitian counterparts. Examples of such novel capabilities include single-mode micro-ring lasers \cite{Feng14Science, Hodaei14Science} that make use of the exceptional points \cite{Brandstetter14NatCommun, Peng14Science,Miao16Science,Feng13NatMaterials} exhibited by systems satisfying parity-time-symmetry, coherent perfect absorption \cite{Wan11Science} in which a thin film of negligible optical losses can be made to completely absorb incident light \cite{Villinger15OL}, and non-reciprocal optical devices that do not require magnetic fields \cite{Peng14NatPhyis, Chang14NatPhoton}. These demonstrations indicate that optical gain and loss can be viewed as resources to be exploited in usefully tuning the response of photonic structures.

We show here that an active planar cavity provided with an appropriate net gain can act as a transparent perfect mirror. Despite the resonant nature of the device, all vestiges of the resonances disappear in the reflected signal at a critical value of pre-lasing gain that we term `Poynting's threshold' \cite{Jahromi16Optica}. Underlying this surprising feature is the emergence at the cavity entrance of a null in Poynting's vector preceding a switch that takes place in the spectral reflectance from depressed resonances to amplified resonances. At this peculiar gain threshold, several salient phenomena occur simultaneously: (a) the structural resonances in reflection disappear; (b) the reflection becomes spectrally flat; (c) light incident on the cavity is -- by necessity -- 100\%-reflected at all wavelengths continuously across the gain bandwidth, independently of the cavity mirrors’ reflectivities; and (d) although the resonances disappear from the reflected signal, they persist in transmission. Therefore, the device at Poynting's threshold serves its dual targeted roles: it cannot be distinguished from a passive perfect mirror when interrogated optically from one side in reflection [Fig.~\ref{fig:Concept}(b)] – while light is nevertheless transmitted and is indeed amplified on resonance [Fig.~\ref{fig:Concept}(c)]. We demonstrate this physical effect here with coherent and incoherent light using active cavities in two configurations: a compact on-chip system and an all-optical-fiber system, and confirm that the cavity reflection resonances all vanish once Poynting's threshold is reached, 100\% reflectivity is achieved continuously across the gain spectrum, and amplified transmission is simultaneously observed.

We describe the concept of a transparent perfect mirror in the context of the non-Hermitian planar-cavity model illustrated in Fig.~\ref{fig:Concept}(d), consisting of an active layer of thickness $L$ with net single-pass power amplification $G$ sandwiched between mirrors $\mathrm{M}_1$ and $\mathrm{M}_2$ of reflectivity $R_1$ and $R_2$, respectively, and an optical probe is incident on $\mathrm{M}_1$ from the left. Here, $G > 1$ (0 dB) corresponds to a net-gain condition. The physical significance of Poynting's threshold can be understood by comparing it to the lasing threshold of the same cavity. The amplification required for lasing is $G_\mathrm{L} = 1/\sqrt{R_1R_2}$, whereupon gain compensates for leakage from both cavity mirrors \cite{Saleh2007}. Poynting's threshold, on the other hand, occurs at the amplification $G_\mathrm{P} = 1/\sqrt{R_2}\leq G_\mathrm{L}$, which suffices to compensate for leakage from only the back (exit) mirror – and thus is a condition that always precedes lasing \cite{Jahromi16Optica}. Poynting's vector along the cavity is $\vec{\mathcal{P}}(z)\approx \{I_\mathrm{f}(z)-I_\mathrm{b}(z)\}\hat{z}$, where $\hat{z}$ is a unit vector in the positive z direction, and $I_\mathrm{f}(z)$ and $I_\mathrm{b}(z)$ are the intensities of the forward and backward propagating waves, respectively, with $I_\mathrm{b}(z)=I_\mathrm{f}(z) R_2 (\frac{G}{G(z)})^2$. (We have neglected a weak interference term contributing to $\vec{\mathcal{P}}(z)$ that is mediated by the complex refractive index of the gain medium \cite{Jahromi16Optica, Dorofeenko12Uspekhi, Ortiz05JOSAA}.) Here $G(z)$ is the amplification experienced after traveling a distance $z$ from $\mathrm{M}_1$; $G(0)=1$ and $G(L)=G$. 

It is clear that $\frac{I_\mathrm{b}(L)}{I_\mathrm{f}(L)}=R_2<1$ such that $\vec{\mathcal{P}}(L)>0$ at the cavity exit $z=L$, a condition that is independent of $G$ and $R_1$, and applies at all wavelengths. In other words, Poynting's vector at the cavity exit is always in the forward direction (that of the incident probe). This condition is solely a consequence of the scattering boundary conditions: the fact that the probe is incident from the left and there is no incoming wave from the right [Fig.~\ref{fig:Concept}(d)]. The transmitted wave is $(1-R_2)I_\mathrm{f}(L)$ for lossless $\mathrm{M}_1$. A different picture emerges at the entrance. By selecting $G=G_\mathrm{P}=1/\sqrt{R_2}$, which we refer to as Poynting's threshold, we have $I_\mathrm{f}(0)=I_\mathrm{b}(0)$ and thus $\vec{\mathcal{P}}(0)=0$: a null develops in Poynting's vector [Fig.~\ref{fig:Concept}(e), middle panel]. This null signifies that the energy flow is extinguished at the entrance plane, which now acts effectively as a perfect reflector to the externally incident probe. That is, when Poynting's vector null coincides with the front edge of the cavity, the reflected signal is equally intense as that of the incident signal. The occurrence of Poynting's threshold is independent of $R_1$, which may be simply the Fresnel reflection from the gain medium.

A simple argument based on optical-energy balance helps elucidate the surprising consequence of this effect. For a lossless mirror $\mathrm{M}_1$, we have $I_{\mathrm{in}}+I_\mathrm{b}(0) = R_{\mathrm{tot}}I_{\mathrm{in}}+I_\mathrm{f}(0)$, where the lefthand-side terms represent the energy of incoming fields at the cavity entrance, while the right-hand terms represent outgoing fields. $R_{\mathrm{tot}}$ is the total reflectivity from the optical structure, and hence incorporates the impact of the structural resonances. However, when $I_\mathrm{b}(0)=I_\mathrm{f}(0)$ at Poynting's threshold $G=G_\mathrm{P}$, we immediately have $R_{\mathrm{tot}}=1$ independently of wavelength or any details of the device construction. Indeed, this unity-reflectivity extends over multiple cavity free-spectral-ranges because satisfying the condition $I_\mathrm{b}(0)=I_\mathrm{f}(0)$ depends on the gain bandwidth and not the cavity linewidth. Simultaneously, since $I_\mathrm{f}(0)$ is non-zero, then $I_\mathrm{f}(L)$ is also non-zero with $I_\mathrm{f}(L)>I_\mathrm{f}(0)$, and an amplified signal thus exits the cavity from $\mathrm{M}_2$.

\begin{figure}[t!]
\begin{center}
\includegraphics[width=8.6cm]{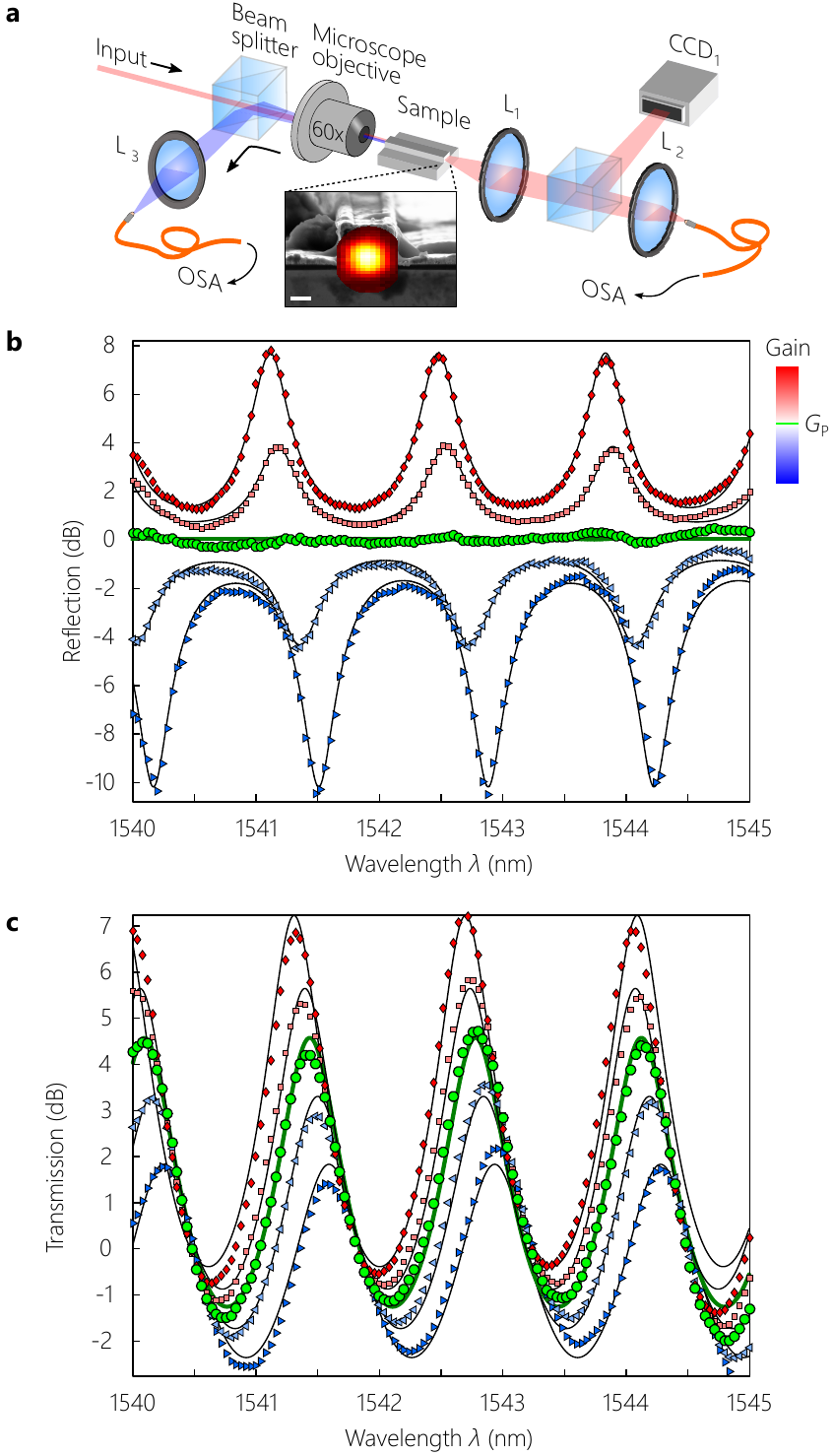}
\end{center}
\caption{Transparent perfect mirror realized using a semiconductor-based active optical waveguide. (a) Schematic of the optical measurement arrangement (Supplement 1, Section S1). The inset is an SEM of the active waveguide facet overlaid with the measured beam profile; scale bar is $1\,\mathrm{\mu m}$. OSA: Optical spectrum analyzer; $\mathrm{L_1}$, $\mathrm{L_2}$, and $\mathrm{L_3}$: lenses of focal length $f=7\,\mathrm{mm}$ (b) Measured spectral reflection $R_\mathrm{tot}(\lambda)$ from and (c) transmission $T_\mathrm{tot}(\lambda)$ through the active waveguide for different values of injected electric current $I$; $R_1=R_2=0.3$, and $L=250\,\mu \mathrm{m}$ The five plots are measurements obtained at the current values $I$ = 8, 10, 12, 14, 17 mA; the lasing threshold in this device is 17.5 mA. Poynting's threshold corresponds to $I=$ 12 mA whereupon the spectral reflection is flat. The measured reflection is shifted by 52 dB such that is shifted by 32.5 dB such that the maximum transmission is 4.5 dB at Poynting's threshold (See Methods).
\label{fig:Waveguide}}
\end{figure}

We have confirmed these predictions in an integrated on-chip active semiconductor waveguide device with electrically driven optical gain in the C-band as illustrated in Fig.~\ref{fig:Waveguide}(a) (Methods and Supplement 1 Section S1). The two waveguide facets provide $R_1=R_2=0.3$, the waveguide length is $L=250$ mm, the free spectral range is $\textrm{FSR}\approx1.38$ nm, and the finesse is $\mathcal{F}=6.36$ (quality factor is $Q\approx3000$; Supplement 1 Section S4). A broadband optical probe is coupled into the waveguide and both the reflected and transmitted spectra are monitored. We plot the spectral reflectance $R_\textrm{tot}(\lambda)$ in Fig.~\ref{fig:Waveguide}(b) and the spectral transmittance $T_\textrm{tot}(\lambda)$ in Fig.~\ref{fig:Waveguide}(c), where $\lambda$ is the wavelength. As we increase the gain $G$, we observe a clear diminishing in the visibility of the reflected resonances until $R_\textrm{tot}(\lambda)$ becomes altogether flat and independent of $\lambda$ at a gain value corresponding to Poynting's threshold. (The carrier-induced spectral shift in the resonant wavelengths with gain is a consequence of the Kramers-Kronig relations \cite{Whalen82JAP, Bennett90JQE}; Methods). On the other hand, $T_\textrm{tot}(\lambda)$ features transmission resonances that increase in magnitude monotonically with gain and persist through Poynting's threshold. The continuous curves in Fig.~\ref{fig:Waveguide}(b,c) are theoretical plots resulting from fitting the data to a model based on analytic expressions for a Fabry-P\'{e}rot resonator (Methods).

One may consider an alternative structure for a perfect transparent mirror: a gain layer followed by a partial mirror such that the gain compensates for the mirror leakage. An anti-reflection coating is required at the front surface of the gain layer to prevent cavity action and the appearance of structural resonances. Nevertheless, even a minuscule reflection due to any imprecision in the anti-reflection coating in conjunction with gain results in clearly visible resonances. To demonstrate this, we have repeated the experiment using an active waveguide with coated facets having reflectivities $R_1=1\%$ and $R_2=90\%$ ($L=1\,\mathrm{mm}$, $\mathrm{FSR}\approx0.35\,\mathrm{nm}$, and $\mathrm{F}=4.55$ or $Q\approx6000$). Despite having R1 at only 1\%, the resonances are quite visible (Supplement 1 Section S2). However, when the gain is increased to reach Poynting's threshold, these resonances vanish, thus yielding once again spectrally flat unity-reflectivity.

\begin{figure}[t!]
\begin{center}
\includegraphics[width=8.6cm]{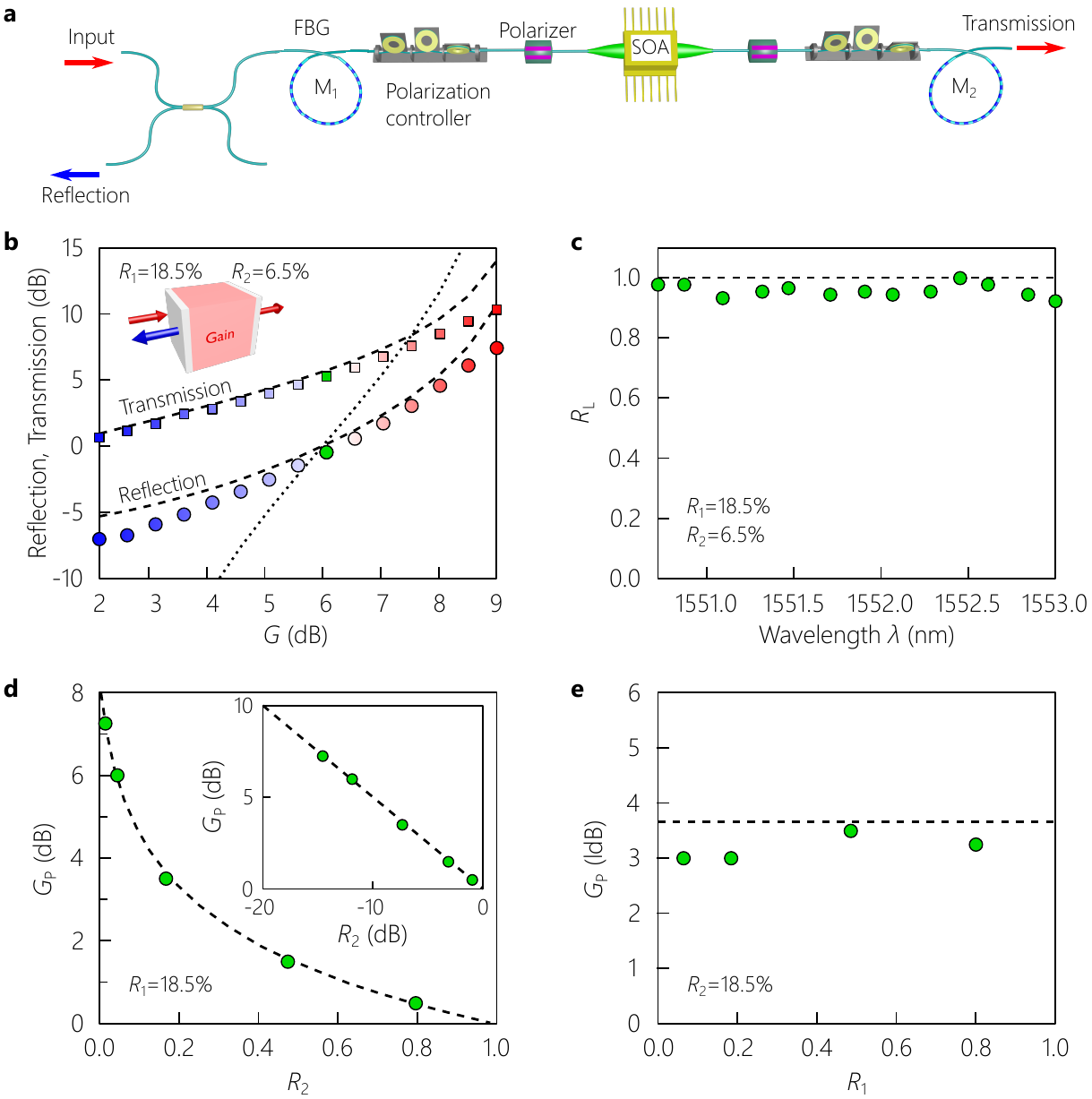}
\end{center}
\caption{(a) Transparent perfect mirror realized in an optical-fiber-based cavity. (a) Schematic depiction of the fiber-based cavity (Supplement 1 Section S3). SOA: semiconductor optical amplifier; FBG: fiber Bragg grating. (b) Measured reflection $R_\mathrm{tot}$ and transmission $T_\mathrm{tot}$ with $G$ at $\lambda=1552\,\mathrm{nm}$ for $R_1=18.5\%$ and $R_2=6.5\%$. Circles ($R_\mathrm{tot}$) and squares ($T_\mathrm{tot}$) are data points. The dashed curves are theoretical predictions for $R_\mathrm{tot}$ and $T_\mathrm{tot}$ after averaging over the cavity FSR, since the tunable optical probe bandwidth is $\ll\,$FSR. The dotted curve is a theoretical prediction for $R_\mathrm{tot}$ strictly on resonance. The two theoretical curves (dashed and dotted) for the reflection intersect at Poynting's threshold $G=G_\mathrm{p}\approx6\,\mathrm{dB}$ (the green colored circular data point), whereupon $R_\mathrm{tot}=1$. The optical probe is a tunable CW laser. (c) Measured spectral reflection at $G=G_\mathrm{p}\approx6\,\mathrm{dB}$ for $R_1=18.5\%$ and $R_2=6.5\%$. (d) Measured Poynting's threshold with $R_2$ at fixed $R_1=18.5\%$ and wavelength $1552\,$nm; the dashed curve is the theoretical prediction $G_\mathrm{P}=1/\sqrt{R_2}$. The inset shows the data on a log-log scale, with the data points lying on a line of slope $-0.5$. (e) Measured $G_\mathrm{P}$ with $R_1$ at fixed $R_2=18.5\%$ and wavelength $1552\,$nm; the dashed curve is the theoretical prediction of a constant $G_\mathrm{P}\approx3.7\,$dB.
\label{fig:Fiber}}
\end{figure}

To confirm our quantitative predictions regarding the behavior of a transparent perfect mirror with respect to its structural parameters, we construct an optical-fiber-based cavity in which all its degrees of freedom are accessible and readily tuned [Fig.~\ref{fig:Fiber}(a)]. We ensure single-mode operation throughout, which allows us to obtain `absolute' values of $R_\textrm{tot}$ without necessitating normalization resulting from coupling efficiencies, as in the first scheme [Fig. ~\ref{fig:Waveguide}]. A semiconductor optical amplifier (SOA) provides broadband gain and $\mathrm{M}_1$ and $\mathrm{M}_2$ are custom-made fiber Bragg gratings with different reflectivity having 5-nm-bandwidth at a central wavelength of $1552\,\mathrm{nm}$. The cavity length is $8\,\mathrm{m}$ corresponding to a FSR of $0.1\,\mathrm{pm}$, and $\mathcal{F}=4.65$ ($Q=2.2\times10^7$). The laser probe has a linewidth of 0.05 nm, corresponding to $\approx500$ FSRs; see Supplement 1 Section S3 for experimental details. In such an arrangement, absolute values of reflectivity and the cavity gain can be measured precisely. First, constructing a cavity with $R_1=18.5\%$ and $R_2=6.5\%$ we confirm that the reflectivity indeed reaches unity at the predicted threshold of $G_\mathrm{P}\approx6\,\mathrm{dB}$ [Fig.~\ref{fig:Fiber}(b)]. Second, we find that this condition extends across most of the cavity bandwidth $\mathtt{\sim}\,2.5\,\mathrm{nm}$ [corresponding to 3000 FSRs; Fig.~\ref{fig:Fiber}(c)]. Third, we determine $G_\mathrm{P}$ while varying $R_2$ for fixed $R_1=18.5\%$, and we obtain the expected $G_\mathrm{P}=1/\sqrt{R_2}$ dependence [Fig.~\ref{fig:Fiber}(d)]. Finally, varying $R_1$ for fixed $R_2=18.5\%$ does not change $G_\mathrm{P}$ [Fig.~\ref{fig:Fiber}(e)].

It is worth noting that amplified spontaneous emission (ASE) is an unavoidable source of noise in any active optical device, which contributes to the measured amplitude in both transmission and reflection. The ratio between the reflected probe and that of the ASE can be defined as the signal-to-noise ratio (S/N) relevant to this configuration, which may be used as a figure-of-merit for the device performance. For example, at Poynting's threshold the average ASE power of the device is -84~dBm according to Fig.~S2a of the Supplementary, where the reflection is -52~dBm from Fig. \ref{fig:Waveguide}b of the main text. This gives $S/N\!=\!32$~dB, implying that the ASE noise is $<0.1\%$ of the measured reflection.

We now address the issue of gain bandwidth in the transparent perfect mirror arrangement. The flatness of the spectral reflection is predicated on the flatness of the gain spectrum. The gain bandwidth of the active semiconductor waveguide is provided in Fig.~S2a of the Supplementary, where there is less than 1~dB variation over a 10-nm bandwidth. The measured gain bandwidth of the fiber-based cavity system is provided in Fig.~2 of Ref. [\cite{Jahromi16Optica}], where these is less than 2~dB variation in gain over a 40-nm bandwidth. In practical settings, one can benefit from well-established methods for \textit{gain equalization} to produce a flat gain spectrum over large bandwidths \cite{Mittelstein89APL,Vengsarkar96OL,Su92PTL,Emori02EL}.

After demonstrating proof-of-principle realizations of a transparent perfect mirror, we consider the prospects of implementing such a concept in planar devices. A material with sufficient optical amplification is required. Currently available OLED materials do not provide high optical gain \cite{Kozlov98JAP}, but two recently developed platforms appear particularly promising. One option is quantum-dot films that provide high net optical gain coefficients $g\approx650$ $\mathrm{cm}^{-1}$; $G=e^{gd}$, where $d$ is the gain medium thickness\cite{Lin16ACSPhotonics}. Such free-standing, chemically and mechanically robust films have been produced over areas on the order of squared-centimeters and thicknesses as high as $500\,\mathrm{nm}$. With these parameters, a transparent perfect mirror with such an active layer requires a back-reflector with $R_2\approx93.7\%$. Another potential platform is perovskites \cite{Green14NatPhoton} from which high-quality films of thickness $1\,\mathrm{mm}$, \cite{Momblona14APLMaterials} and gain coefficient $g=250$ $\mathrm{cm}^{-1}$, \cite{Xing14NatMaterials} have been reported, which requires $R_2\approx95.6\%$. Unwanted or spurious reflections from transmitted light can be eliminated using an isolator or polarization optics behind the mirror \cite{Bengoechea2004}. To avoid lasing in practical realizations, $\mathrm{M}_1$ can be selected to be a low-reflectivity mirror, perhaps simply the Fresnel reflection from the gain medium. Both suggested materials provide large broadband optical gain. Specifically in the case of QD-doped polymer films, multiple QD species can be combined in the same film to produce broadband gain since we require only inhomogeneous broadening.

Two limitations of transparent perfect mirrors must be pointed out. First, in realizing this effect in a planar geometry, Poynting's threshold becomes dependent on polarization and incidence angle. We find that the transparent-perfect-mirror effect, however, degrades gracefully: considering an angular bandwidth of $20^\circ$ centered on normal incidence, a maximal reduction or increase in reflectivity of only $\mathtt{\sim}\,6.5\%$ is observed (Supplement 1 Section S5). Second, although flat spectral reflectivity is maintained over the gain bandwidth, observing the reflection of coherent pulses may reveal the cavity structure via its decay time. The effect is related to the transit time $\tau\,\mathtt{\sim}\,nL/c$ across the cavity. For the device used in Fig.~\ref{fig:Waveguide} we have $\tau\approx3\,\mathrm{ps}$. Therefore, an incident pulse whose width is on the order of $3\,\mathrm{ps}$ is deformed and distinct echoes can be observed in the time domain in the reflected signal at Poynting's threshold. An incident data stream at a rate of less than 100 Gb/s can be reflected without noticeable distortion. In the case of the planar realization suggested above where $L$ $\mathtt{\sim}\,1\,\mathrm{\mu m}$, we have $\tau\,\mathtt{\sim}\,15\,\mathrm{fs}$, thereby increasing the resources required to uncover the underlying structure of the perfect mirror in the time domain (see Supplement 1 Section S6 for a full analysis).

We have shown that an appropriately designed active optical cavity can be indistinguishable from a passive perfect mirror, but nevertheless transmits light. This compact single-port device can be used in camouflage and applications that demand privacy and security. Furthermore, this result hints at opportunities for invisibility by molding the flow of light in non-Hermitian structures that contain optical gain. Finally, since the device relies on coherent amplification, it would be interesting to investigate its impact on the incident photon statistics \cite{Kondakci15NatPhys}, especially when non-classical states of light are utilized. More generally, the concept of Poynting's threshold is a universal wave phenomenon that can be exploited to create transparent perfect mirrors for microwaves, electronics, acoustics, phononics, and electron beams.

\section{Methods}
\textbf{Chip Manufacturing.} The chip manufacturing starts with epitaxial growth of the layered structure on an InP substrate \cite{Savolainen99JEM}. We make use of the AlGaInAsP material system and design a separate confinement heterostructure with a multi-quantum well (MQW) active region. A ridge waveguide was manufactured using standard lithographic methods to define a 2-$\mathrm{\mu}$m-wide ridge and etching through most of the upper cladding layer to ensure operation in a single spatial mode. Proper isolation and top electrical contacts \cite{Vilokkinen04EL} are fabricated followed by wafer-thinning and back-side metallization for the cathode contact. The processed material is then cleaved into bars with 250-$\mathrm{\mu}$m cavity length. No optical coatings were applied on the facets, which are left as-cleaved. Finally, the bars are separated into 300-$\mathrm{\mu}$m-wide chips mounted on C-mount packages that act as heat sinks and provide simple electrical connections and mechanical mounting.

\textbf{Carrier-Induced Resonance Shift.} The waveguide is fabricated on an InP substrate, with InGaAsP/AlGaInAs quantum-wells as the active MQW layers. It has been shown experimentally that carrier injection in InP, which provides optical gain, modifies the imaginary part of the refractive index [\cite{Whalen82JAP, Stone82APL,Chusseau98APL}]. Consequently, current injection produces a change in the real part of the refractive index of the material per the Kramers-Kronig relation. This effect is attributed to three main carrier effects: free-carrier absorption, bandgap shrinkage, and conduction band filling \cite{Bennett90JQE}. Previous results show that the index of InP or $\mathrm{In}_{0.82}\mathrm{Ga}_{0.18}\mathrm{As}_{0.40}\mathrm{P}_{0.60}$ initially decreases with induced carrier concentration before reaching a local minimum, after which the refractive index increases with further increase in the induced carrier concentration. Using the data presented in Fig.~\ref{fig:Waveguide}(b) (in addition to measurements at other currents extending above the lasing threshold), we obtain the change in index $\Delta n$ with wavelength. The resonance condition is characterized by $k_{0}nL=2m\pi$, where $k_{0}=\frac{2\pi}{\lambda}$, $\lambda$ is the free-space wavelength, $m$ is integer, and $n$ is the real part of the refractive index (we have ignored here the phase due to reflection at the waveguide facets). From this we have $\frac{\Delta n}{n}=\frac{\Delta \lambda}{\lambda}$, which enables us to plot the results shown in Supplement 1 Fig. S4.

\textbf{Data-Fitting Procedure.}
We present here the data-fitting methodology used in Fig.~\ref{fig:Waveguide}(b,c). We first obtain an analytic expression for reflection from the symmetric Fabry-P\'{e}rot  waveguide using the transfer-matrix method,
\begin{equation}\label{eq:TotalReflection}
R_\textrm{tot} = 1- \frac{(1-R)(1-TG^2)}{1+R^2\,G^2-2RG\,\textrm{cos}\varphi},
\end{equation}
where $T=1-R$, $\varphi=\alpha_1+\alpha_2+2nkL$, $k=\frac{2\pi}{\lambda}$, $\lambda$ is the free-spacewavelength, $L$ is the waveguide length, $\alpha_1$ anda $\alpha_2$ are the phases incurred upon reflection from the two facets. The expression can be recast as follows:
\begin{equation}\label{eq:ParametrizedReflection}
R_\textrm{tot} = 1-\frac{(1-R)(1-RG^2)/(1-RG)^2}{1+\Big(\frac{2\sqrt{RG}}{1-RG}\, \mathrm{sin}(\frac{4\pi}{\lambda}nL+\alpha_1+\alpha_2)\Big)^2}\, \longrightarrow \, \Bigg(1-\frac{p_1}{1+\Big(p_2\, \mathrm{sin}(p_3/\lambda+p_4)\Big)^2}\Bigg)\frac{1}{p_5},
\end{equation}
where we have introduced fitting parameters $p_j$ (Supplement 1 Table S1). The last parameter is $p_5$ introduced as an overall normalization factor. Not all fitting parameters are independent. For example, $p_1$ and $p_2$ are readily related through equation \ref{eq:ParametrizedReflection}, from which we obtain $(1-R)(1-RG^2)=p_1(1-RG)^2$.

The procedure for fitting the transmission data $T_\mathrm{tot}$ in Fig.~\ref{fig:Waveguide}(c) is similar. The model is based on an analytic expression for $T_\mathrm{tot}$ (see Supplement 1 Table S2 for the fitting parameters):
\begin{equation}\label{eq:Transmission}
\begin{aligned}
T_\textrm{tot} = \frac{(1-R)^2}{(1+R_1R_2\,G^2)-2G\sqrt{R_1R_2}\,\textrm{cos}\varphi}=\frac{(1-R)^2/(1-RG)^2}{1+\Big(\frac{2\sqrt{RG}}{1-RG}\, \mathrm{sin}(\frac{4\pi}{\lambda}nL+\alpha_1+\alpha_2)\Big)^2}\, \\ \longrightarrow \, \Bigg(1-\frac{p_1}{1+\Big(p_2\, \mathrm{sin}(p_3/\lambda+p_4)\Big)^2}\Bigg)\frac{1}{p_5}.
\end{aligned}
\end{equation}

\section{Funding}
US Air Force Office of Scientific Research (AFOSR) under contract AFOSR MURI contract FA9550-14-1-0037.

\begin{acknowledgement}
We gratefully thank Luke Lester, Demetrios N. Christodoulides, Kyle Renshaw, Guifang Li, Bin Huang, He Wen, Mina Bayat, Parinaz Aleahmad, Felix Tan, Himansu Pattanaik, and Imen Rezadad for helpful discussions and loan of equipment. We thank Joshua J. Kaufman for assistance with scanning electron microscopy and Kristi Peters for help preparing the illustrations.
\end{acknowledgement}

\begin{suppinfo}

\end{suppinfo}


\bibliography{References.bib}

\end{document}